\documentclass[10pt]{iopart}
\usepackage{floatflt}
\usepackage{wrapfig}
\usepackage{graphicx}
\begin{document}

\def\PRL{Phys. Rev. Lett. }
\def\PRC{Phys. Rev. C }
\def\PRD{Phys. Rev. D }
\def\PLB{Phys. Lett. B }
\def\NPA{Nucl. Phys. A }
\def\pT{\mbox{$p_T$}}
\def\pt{\mbox{$p_T$}}
\def\v2{\mbox{$v_2$}}
\def\sqrtsNN{\mbox{$\sqrt{s_{NN}}$}}
\def\jt{\mbox{$j_T$}}
\def\kt{\mbox{$k_T$}}
\def\et{\mbox{$E_T$}}
\def\mkt{\mbox{$\langle k_T\rangle$}}
\def\met{\mbox{$\langle E_T\rangle$}}
\def\mjt{\mbox{$\langle j_T\rangle$}}

\title[Jets as a Probe of Dense Matter at RHIC]{Jets as a Probe of Dense Matter at RHIC}

\author{Kirill Filimonov}

\address{\dag\ Nuclear Science Division, Lawrence Berkeley National Laboratory, 1 Cyclotron Road, Berkeley CA 94720}
\ead{KVFilimonov@lbl.gov}

\begin{abstract}
Jet quenching in the matter created in high energy 
nucleus-nucleus collisions provides a tomographic tool to probe
the medium properties. 
Recent experimental results on jet production at 
the Relativistic Heavy-Ion Collider (RHIC) are reviewed. 
Jet properties in p+p and d+Au collisions have been measured,
establishing 
the baseline for studying jet modification in heavy-ion collisions.
Current progress on detailed studies of high transverse momentum
production in Au+Au collisions is discussed, with an emphasis on 
dihadron correlation measurements.
\end{abstract}




Jet quenching, first predicted \cite{quenching}, and then observed at RHIC 
\cite{suppression}, 
is one of the most significant results
of the heavy-ion collider program so far. 
Attenuation of jets enables tomographic analysis of the
created matter  using the jet as an 
effectively external probe of the medium \cite{tomography}. 
To gain insight into the 
detailed mechanisms of the interaction of the fast propagating parton
with the medium and, ultimately, to extract the medium properties,
jet production and its
modification must be studied in a multi-parameter space 
as a function 
of \pt, particle species/flavor, centrality, and reaction plane.
It is vital to perform the same measurements both in Au+Au and simpler
reference systems such as p+p and d+Au, to calibrate the limited set of jet-related
observables available in the complex Au+Au environment.

Three different phenomena related to 
partonic energy loss have been observed 
for particle production at high \pt~ in Au+Au collisions at
RHIC:
\begin{itemize}
\item{strong suppression 
of inclusive hadron spectra in central collisions \cite{suppression}}
\item{large azimuthal anisotropy in non-central collisions \cite{flow}}
\item{disappearance of back-to-back azimuthal correlations in central collisions \cite{btob}} 
\end{itemize}

Between the last Quark Matter conference \cite{qm02} and this one, 
critical high \pt~measurements were performed with d+Au collisions \cite{dAuPhobos,dAuPhenix,dAuStar,dAuBrahms}. 
This control experiment distinguished 
cold and hot nuclear matter effects and the results confirmed that 
nuclear attenuation observed in central Au+Au collisions
is due to final-state interactions of jets in the dense matter formed
in heavy-ion collisions. Experimental results on inclusive
spectra and azimuthal correlations obtained prior to this conference 
are reviewed by D.~d'Enterria \cite{denterria} and M.~Miller \cite{miller}.
The quantitative applications of the theory and phenomenology 
of medium-induced
energy loss are summarized by I.~Vitev \cite{vitev}.

At this conference, qualitatively new and important results
were presented from the first full reconstruction of jets at RHIC.
Full jet reconstruction in p+p and d+Au collisions  
calibrates the dihadron observables used to infer jet properties in
Au+Au collisions.   
New, in-depth, analyses of the older Au+Au dataset were also reported,
in particular in the area of dihadron correlations.
Whereas measurements of single hadron
distributions cover \pt~up to 12 GeV/c, 
hadron-hadron correlation
studies are currently statistics limited to rather moderate \pt$\sim$2-6 GeV/c.
In this region of transverse momentum, an enhancement of baryon to meson 
ratios has been observed \cite{baryonmeson}, indicating modifications to the 
particle production mechanism compared to p+p collisions and 
pQCD calculations.
Nevertheless,
some intriguing new results have emerged: 
direct evidence that back-to-back dijet quenching depends on the 
azimuthal orientation of the jets relative to the reaction plane;
jet-like correlations are shown to exist for trigger 
mesons, baryons, $\Lambda$ and $K^0$; 
away-side azimuthal distributions 
associated with 
a high \pt~trigger hadron 
are suggestive of statistical momentum balance, perhaps indicating 
equilibration of the medium-induced gluon radiation.
Below I will review these and other experimental results presented at
this conference.

\section{Jet Properties in p+p and d+Au}
Results on first full jet reconstruction at RHIC have been presented
by STAR \cite{henry}. Inclusive jets have been directly measured
in p+p and d+Au collisions using calorimetry and charged
particle tracking. 
Full jet reconstruction reduces the uncertainties due to fragmentation of partons into hadrons that are present in di-hadron correlation studies.
 
\begin{figure}[b]
\begin{center}{\includegraphics*[%
  keepaspectratio,
  width=0.48\columnwidth]{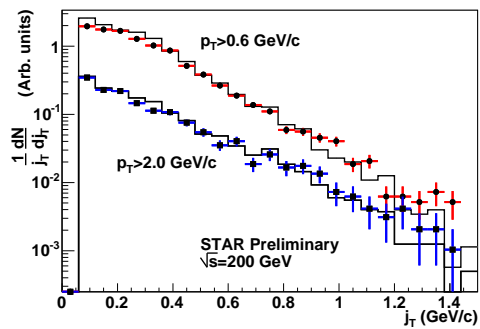}}
\end{center}
\vspace*{-0.05in}
\caption{
Jet $j_T$ distributions in p+p collisions from STAR \cite{henry}
for two hadron $p_T$ thresholds. Histograms are Pythia Monte Carlo events.
\label{fig:jt}}
\end{figure}

Non-perturbative fragmentation processes result in an approximately 
Gaussian jet cone. The transverse shape of a jet is 
characterized by \jt=$p_{hadron}\sin\theta$, 
the hadron momentum component
perpendicular to the jet thrust axis. 
Figure~\ref{fig:jt} shows that 
measured \jt~distributions in p+p
collisions agree well with Pythia which accurately describes jet features at other 
energies.
For jets with \met$\sim$11 GeV, STAR measured \mjt=515$\pm$50 MeV/c.
PHENIX \cite{rak} performed di-hadron 
correlation measurements of the near-angle azimuthal ($\Delta\phi$)
distributions and extracted \mjt=510$\pm$10 MeV/c  
which agrees well with the STAR value obtained using full jet reconstruction.

Experiments also reported the measurements of \kt, or the transverse
momentum of an individual parton within a nucleon (\kt$_{pp}$) or
nucleus (\kt$_{nucl}$), inferred from the acoplanarity
in the dijet production (away-side $\Delta\phi$ distribution).
Non-zero values of \kt~in hadronic collisions are usually attributed to 
the intrinsic transverse momentum of the initial state partons
due to the finite size of the incoming hadron and to multiple soft-gluon emission by the partons prior to the
hard scattering.
Measurements of dimuon, diphoton and dijet production in hadronic collisions
have indicated the 
presence of significant effective \kt~\cite{apanasevich}.

\begin{figure}[ht]
\begin{flushleft}{\includegraphics*[%
  keepaspectratio,
  width=0.48\columnwidth]{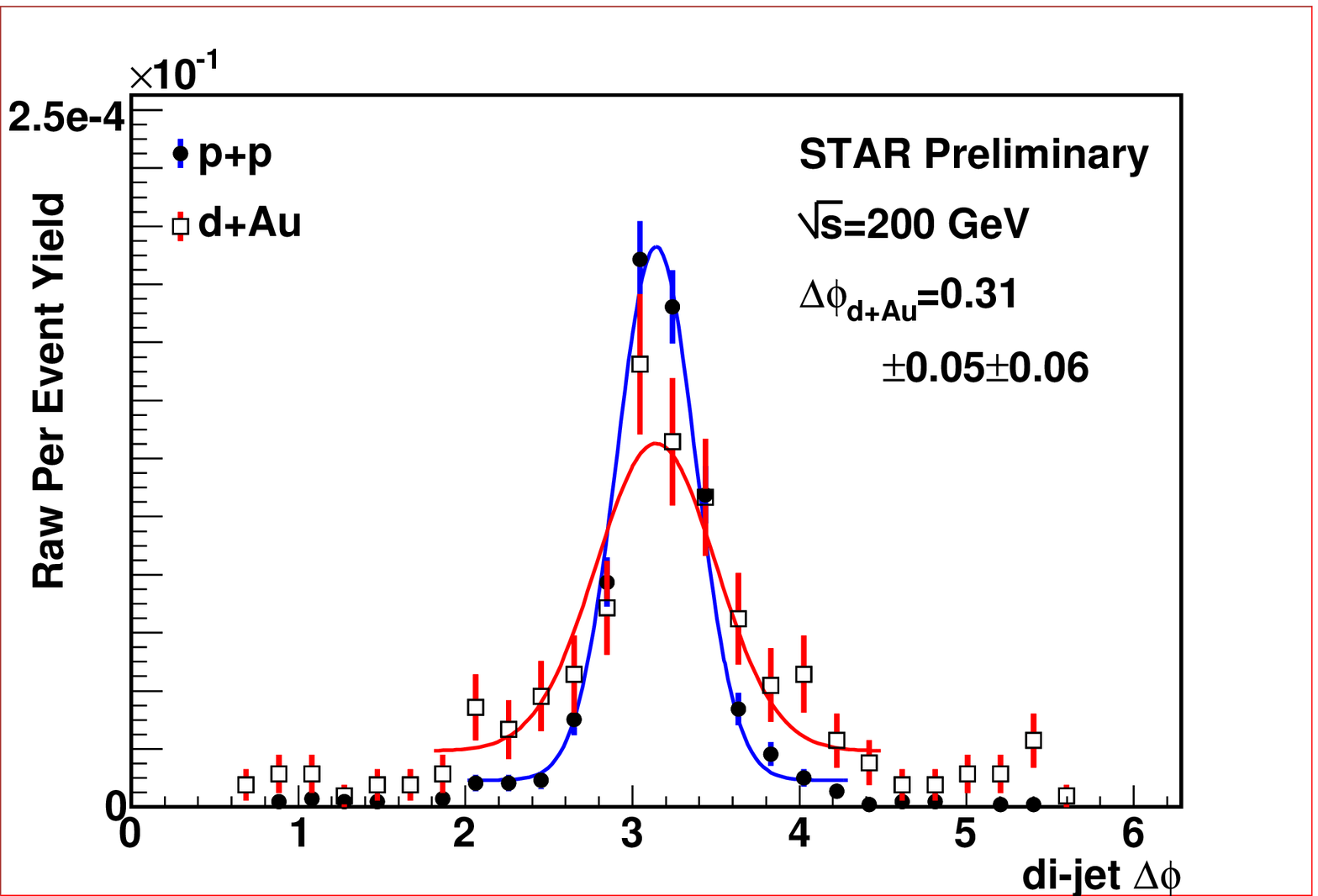}}\hfill{\includegraphics*[%
  height=1.9in,
  width=0.48\columnwidth]{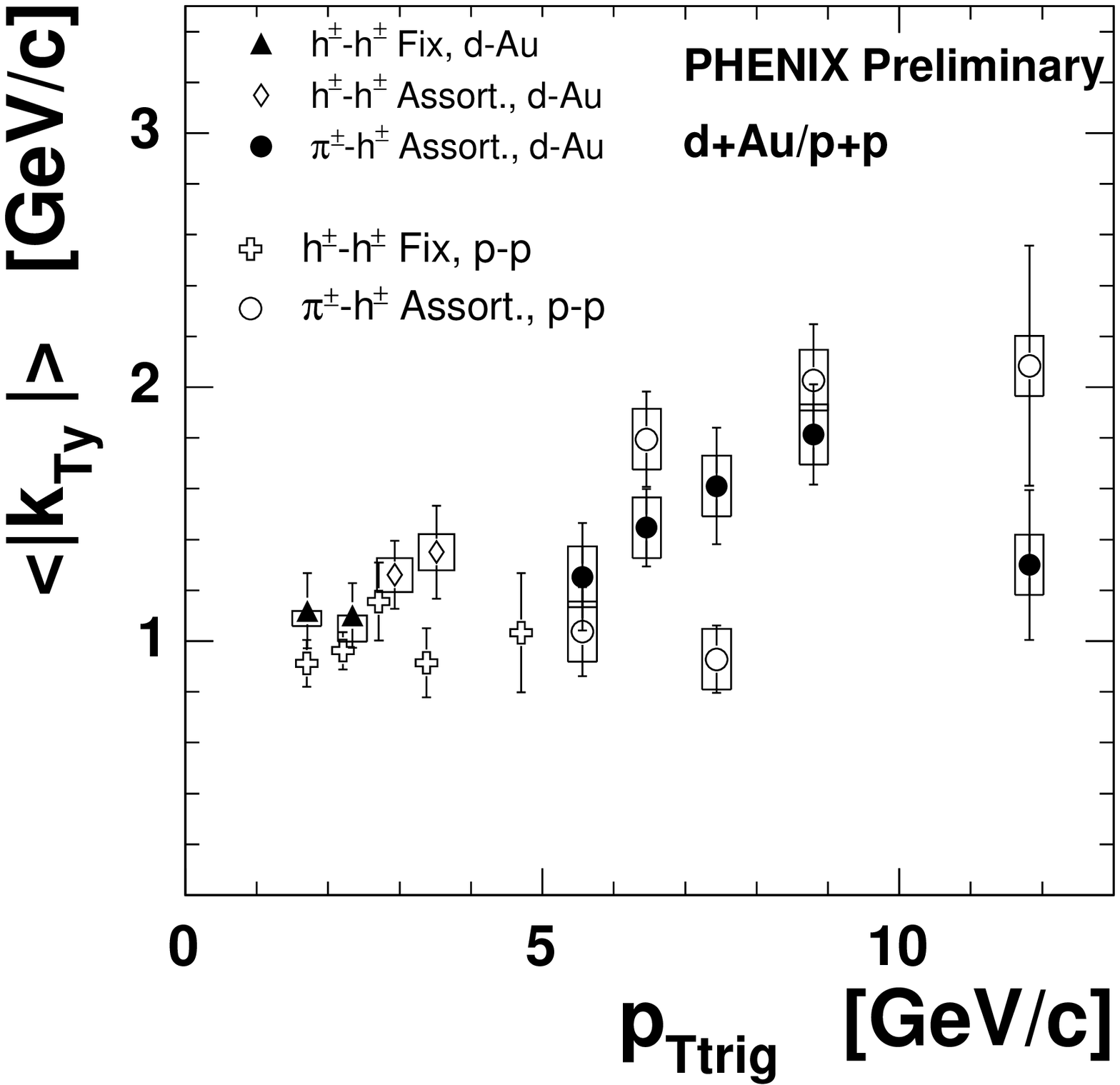}}\end{flushleft}
\vspace*{-0.05in}
\caption{Left: Di-jet  $\Delta\phi$ distribution in p+p
and d+Au collisions at $\sqrt{s}$=200 GeV from STAR \cite{henry}.
Right: Variation of $\langle|k_{Ty}|\rangle$ with \pt$_{trig}$ measured by
PHENIX \cite{rak}. 
\label{fig:kt}}
\end{figure}

Figure~\ref{fig:kt} (left) shows the dijet $\Delta\phi$ distribution in p+p
and d+Au collisions measured by STAR at $\sqrt{s}$=200 GeV,
for jets of energy on the order of \met$\sim$ 13 GeV.
Larger width of the away-side peak is found in d+Au collisions.  
Taking \kt$^2_{dAu}$=\kt$^2_{pp}$+\kt$^2_{nucl}$ indicates
a finite value for \kt$_{nucl}$. 
PHENIX \cite{rak} reported systematic
measurements of di-hadron correlations in p+p and d+Au collisions
as a function of trigger and associated hadron \pt~(Fig.~\ref{fig:kt}, right). 
\mkt~shows a tendency to increase with the larger trigger \pt,
with no significant difference in \kt-values measured
in p+p and d+Au collisions.
STAR quotes the preliminary value of the intrinsic RMS \kt$_{pp}$=2.3$\pm$0.4$\pm^{0.67}_{1.1}$ GeV/c and RMS \kt$_{nucl}$=2.8$\pm$1.2$\pm$1.0 GeV/c.
The RMS value of the two-dimensional per-parton vector
${\bf k}_T$ used by STAR is related to  $\langle|k_{Ty}|\rangle$ reported
by PHENIX
as $\sqrt{\langle k_T^2\rangle}=\sqrt{\pi}\langle|k_{Ty}|\rangle$. 
Whereas STAR and PHENIX measurements of \kt$_{pp}$ are consistent with each other and the world data systematics as a function of $\sqrt{s}$, 
within current statistical and systematic uncertainties
it is too early to conclude about possible discrepancies 
in the measurements of \kt$_{nucl}$.

PHENIX \cite{rak} also extracted \kt~from the di-hadron azimuthal distributions in  Au+Au data for $p_T^{\rm trig}$=2.5-4.0 GeV/c and $p_T^{\rm assoc}$=1.0-2.5 GeV/c. They report 
a dramatic increase of \mkt~with centrality. Interpretation of this
result, however, is not straightforward. Strong suppression 
of away-side jets and large values of elliptic flow measured at RHIC
complicate the  analysis of di-hadron azimuthal correlations in Au+Au data.
Higher statistics and larger \pt-scales
are needed to effectively decouple the jet and elliptic flow contributions
to the di-hadron distributions.

\section{Inclusive Spectra}

The \pt-range of single inclusive hadron spectra measured  
at RHIC has become truly remarkable, reaching \pt=10-15 GeV/c.
Since the bulk of the data on inclusive spectra in p+p, d+Au and Au+Au 
collisions has already been published, I will not review these results.
Significant new measurements of hard photons and 
heavy quarks
have been presented at this conference. Direct photon (PHENIX) 
and open charm (STAR) measurements in d+Au collisions extend up to 
\pt=11 GeV/c and are reviewed by R. Averbeck \cite{averbeck}.

\begin{figure}[ht]
\begin{center}
\includegraphics[width=100mm,height=60mm]{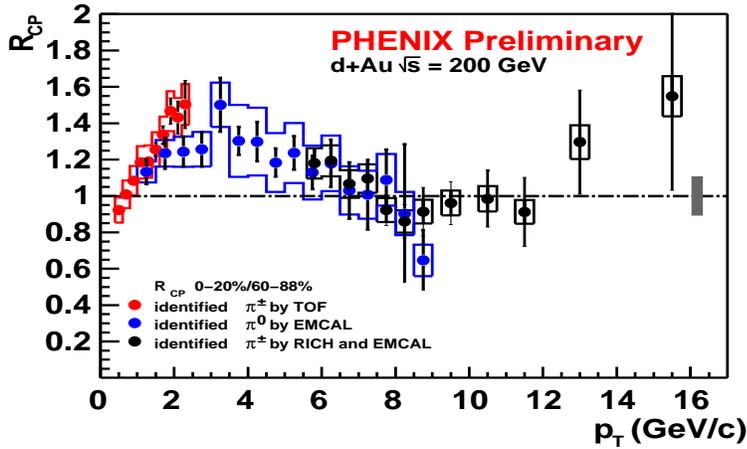}
\caption{PHENIX data \cite{bosing} 
on nuclear modification factor in d+Au collisions at $\sqrt{s}$=200 GeV.
\label{fig:spectra}}
\end{center}
\end{figure}

One of the important new results regarding spectra was shown by PHENIX 
\cite{bosing}. New data on the nuclear modification
factor, the ratio of central and peripheral d+Au
spectra normalized by the number of binary collisions, 
are shown in Fig.~\ref{fig:spectra}. 
It was previously observed that in d+Au collisions, 
nuclear effects are significant (nuclear modification factor exceeds unity, 
manifestation of ``Cronin'' enhancement) in the region of \pt=2-7 GeV/c 
\cite{dAuPhenix,dAuStar}.
The new PHENIX data extend up to \pt=12-15 GeV/c
and for \pt$>$6-8 GeV/c the yields scale with the number of binary collisions,
consistent with no significant initial state nuclear 
effects for high transverse momentum hadrons produced at midrapidity.

PHENIX also presented detailed centrality dependence of the nuclear 
modification factor in  d+Au and Au+Au collisions for charged hadrons
and neutral pions, 
with the latter extending to 15 GeV/c in Au+Au \cite{bosing}, as well as for $\pi^++\pi^-$ and $p+\bar{p}$
\cite{matathias}. In peripheral collisions the particle yields
are consistent with binary scaling for both colliding 
systems. Going from peripheral to central collisions, 
progressive enhancement of particle production in d+Au collisions
is seen, in contrast to stronger suppression for Au+Au collisions.
This strong centrality dependence in d+Au differs, however, from the 
STAR observations of weak to no centrality dependence of binary-scaled
hadron yields \cite{dAuStar,dAuStarId}.
The flavor dependence of the Cronin effect in d+Au collisions is shown to be
not large enough to explain the $p/\pi$ ratio measured in Au+Au collisions.
STAR data \cite{lamont} on $\Lambda$, $K^0$, $\Xi$ and $K^*(892)$
production in Au+Au collisions
demonstrate that the difference is due to a baryon/meson effect and 
seems to be confined to the region of \pt=2-6 GeV/c.

\section{Chemistry of a Jet}

Identified hadron jet studies in Au+Au collisions 
may be able to probe
the energy loss difference of quarks versus
gluons as well as possible modifications in flavor composition and 
multiplicities in a jet due to coalescence/recombination 
in the medium.
Several aspects of particle production in central Au+Au collisions at RHIC
in the \pt-range of 2-6 GeV/c are incompatible with
jet fragmentation in simpler systems: large $p/\pi$ ratio \cite{matathias},
antibaryon to baryon ratio constant with \pt~\cite{lamont}, 
different 
suppression of proton/pion \cite{matathias} 
and lambda/kaon yields \cite{lamont},
and deviation from hydrodynamic mass-ordering in strength
of elliptic flow of pions/protons \cite{idflowphenix} and kaons/lambdas \cite{idflowstar}.
Models based on coalescence of thermal partons are 
successful in qualitatively
describing the features observed in the data \cite{fries}.

\begin{figure}[ht]
\vspace{-0.1cm}
\begin{center}
\resizebox{
\textwidth}{!}{
\includegraphics{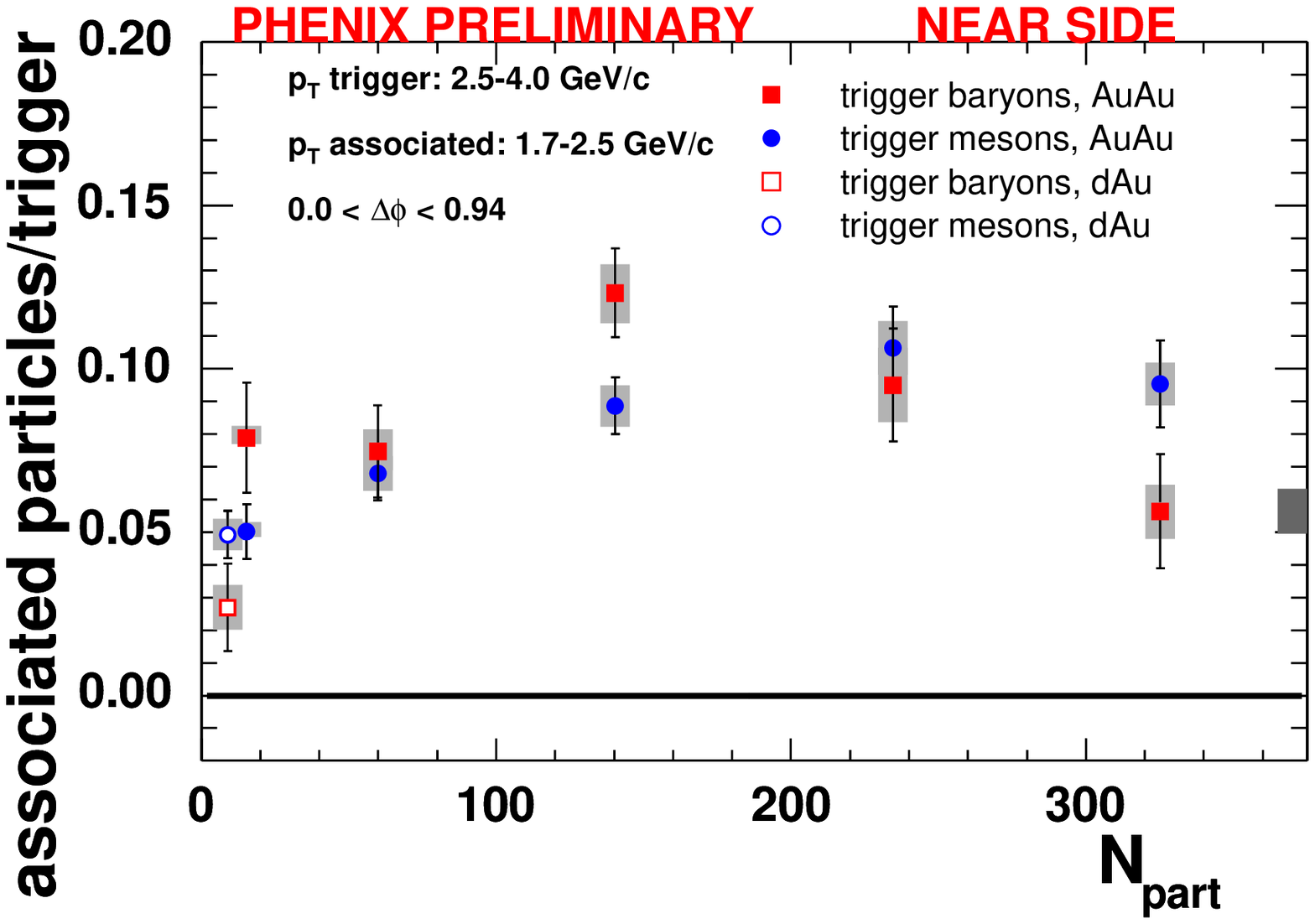}
\includegraphics{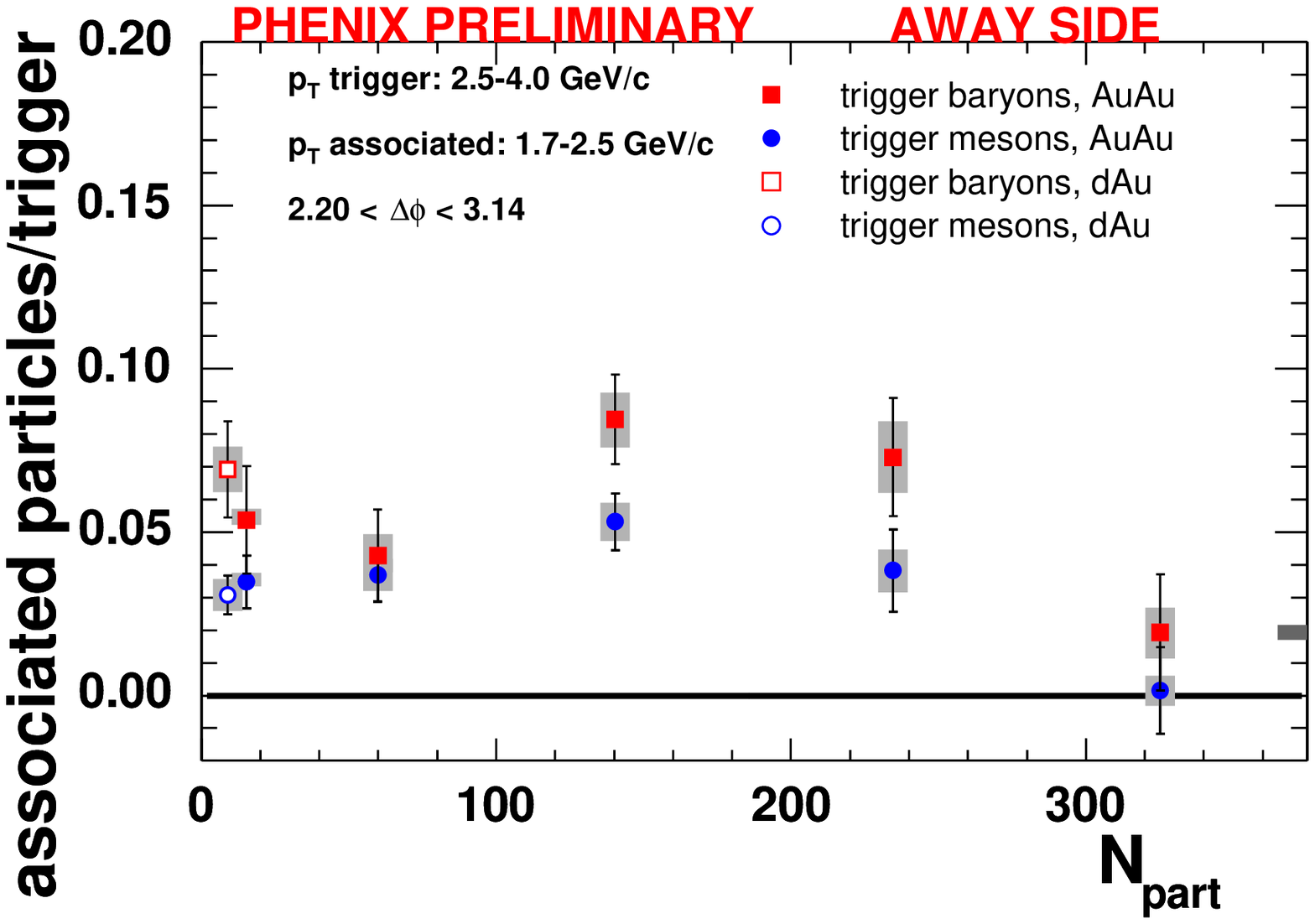}}
\vspace{-0.5cm}
\caption{PHENIX data \cite{sickles} on near (left) and away-side (right) associated particle yields per
baryon/meson trigger.
\label{fig:barmes}}
\end{center}
\vspace{-0.2cm}
\end{figure}

On the other hand, new data
 on the relative azimuthal distributions of identified 
baryons and mesons (PHENIX \cite{sickles}), $\Lambda$ and $K^0$ 
(STAR \cite{guo}) with $p_T>$2.5 GeV/c and charged 
particles show finite jet-like correlations.
STAR finds that $\Lambda$ and $\bar{\Lambda}$ correlations with charged
hadrons in central Au+Au collisions 
have different trigger \pt~dependences.
PHENIX results on the centrality dependence of 
associated particle yields per identified
trigger are shown in Fig.~\ref{fig:barmes}.
No significant change in yield on the near-side, 
or decrease expected in a naive 
coalescence picture, is observed 
as a function of the number of participants. On the away-side,
an overall decrease in the yields is observed similar to 
back-to-back suppression of charged particles \cite{btob}.
The data perhaps indicate slightly larger yields on the away side
for the baryon triggers. The tendencies shown in the data are quite
indecisive at this point; significantly
higher \pt~probes are required to clarify the observations. 
It will be interesting to apply asymmetric \pt~thresholds
for baryons and mesons (with the ratio corresponding to the number of
constituent quarks) to test coalescence models.
Theoretical progress is also needed to understand whether 
coalescence/recombination
may result in sizable correlations in the azimuthal distributions \cite{fries}.

\section{Jet Production versus Reaction Plane}
\begin{figure}[ht]
\begin{center}
\includegraphics[width=90mm]{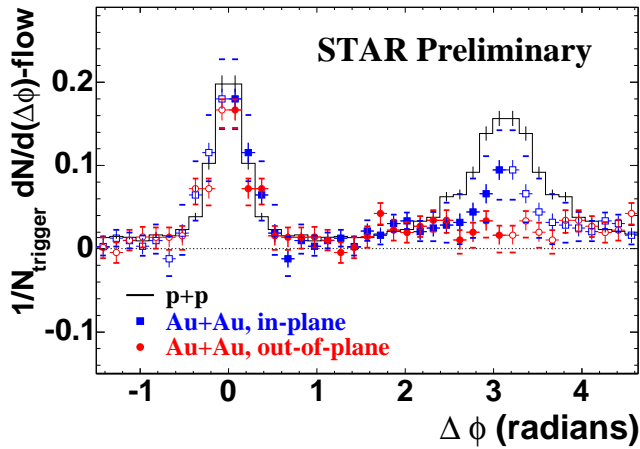}
\caption{STAR data \cite{aihong} 
on azimuthal distributions of associated particles
 for trigger particles in-plane (squares) and out-of-plane (circles)  
for Au+Au collisions at centrality 20\%-60\%, compared with p+p reference data 
(histogram). 
\label{fig:inOut}}
\end{center}
\end{figure}

Direct evidence of the variation of the strength of back-to-back 
suppression  with azimuthal orientation of the jets relative
to the reaction plane was presented by STAR \cite{aihong}.
Di-hadron correlations were measured  
in non-central Au+Au collisions at \sqrtsNN= 200~GeV for 
trigger particles with \pt=4-6 GeV/c sorted 
in the direction of the reaction plane angle (in-plane)
and perpendicular to it (out-of-plane).
The trigger particles were paired with associated particles with 
2~GeV$/c < p_T < p_T^{\rm trig}$.
Figure~\ref{fig:inOut} shows the azimuthal distribution of associated 
particles in Au+Au (elliptic flow subtracted \cite{inout}) 
compared with p+p reference data. 
The near-side 
jet-like correlations measured in Au+Au are 
similar to those measured in p+p collisions. The back-to-back 
correlations measured in Au+Au 
collisions for in-plane trigger particles are suppressed compared 
to p+p, and even more suppressed for the out-of-plane trigger particles.  
Such behavior is naturally predicted by jet quenching 
models, where the energy  
loss of a parton depends on the distance traveled through the 
dense medium \cite{wang}.
Due to energy loss in these models, the high \pt~trigger biases the initial production point to be near 
the surface so the near-side correlations should be similar to those
seen in p+p collisions. The away-side correlations are more suppressed
when the trigger hadron is emitted perpendicular to the reaction plane.
Measurements of the effect as a function of centrality will allow to 
precisely determine the energy loss dependence on the path length.

Results on single particle high $p_T$ charged hadron (STAR \cite{aihong}) 
and $\pi^0$ production (PHENIX \cite{kaneta})
with respect to the reaction plane 
were also presented. Azimuthal anisotropy, quantified by the second harmonic 
Fourier coefficient $v_2$, is found to reach its
maximum at $p_T\sim$3~GeV, decreasing very slowly if at all
up to $p_T\sim$7-10 GeV/c. STAR performed 
higher order cumulant analysis for $v_2$ and 
compared azimuthal correlations measured in p+p collisions to
those in Au+Au.  The results confirm 
strong elliptic flow in mid-central Au+Au
collisions at least up to $p_T\sim7$~GeV/c, qualitatively
consistent with jet quenching. The strength of the $v_2$ signal remains
too large to be explained by quenching alone \cite{shuryak}.
Measurements of $v_2$ that are free of non-flow 
effects in the \pt-region which is clearly dominated by jets 
(most likely, \pt$>7$ GeV/c) are highly desirable.

\section{Reconstruction of the Lost Energy}

\begin{figure}[ht]
\begin{center}
\begin{flushleft}{\includegraphics*[%
  keepaspectratio,
  width=0.48\columnwidth]{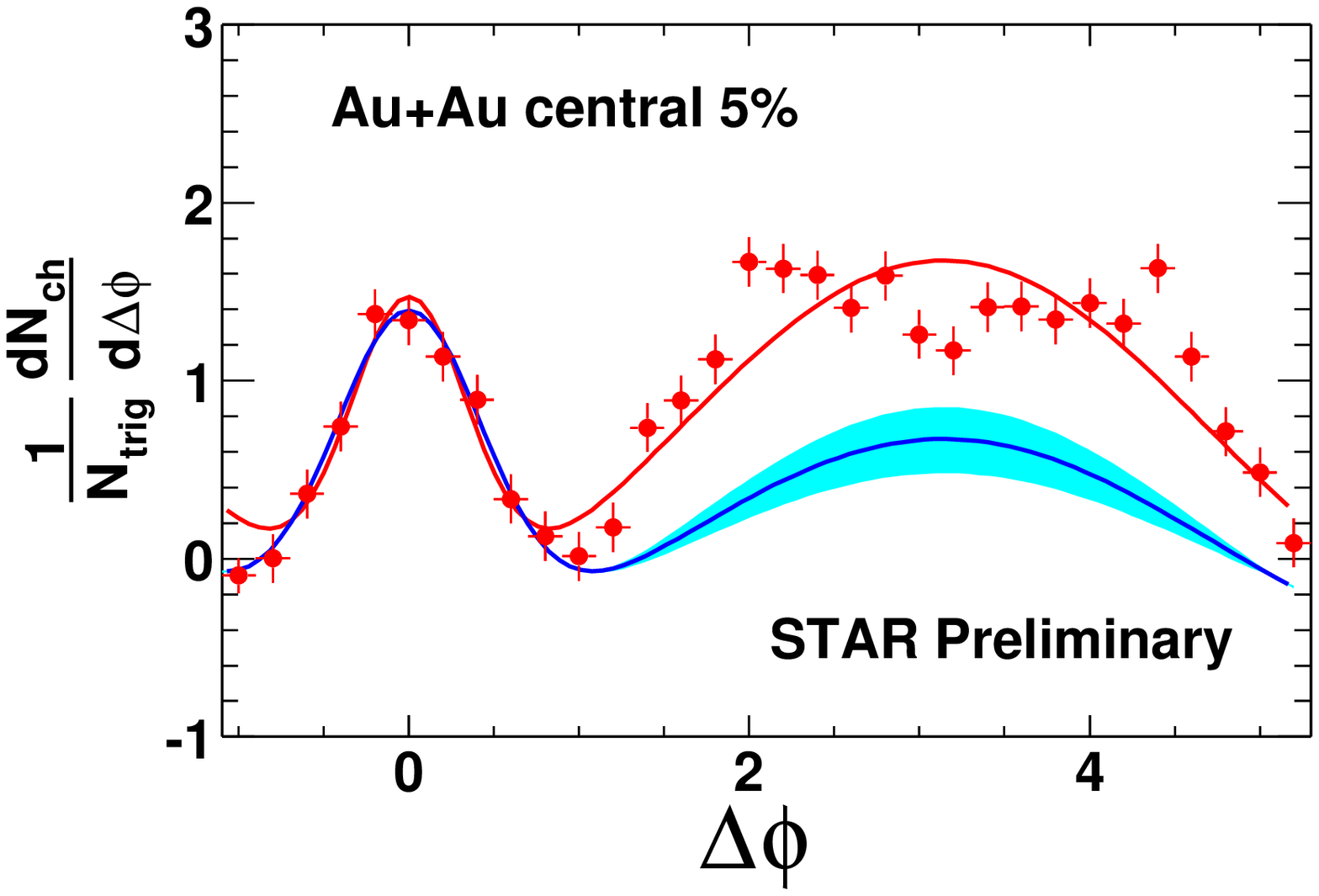}}\hfill{\includegraphics*[%
  height=2.1in,
  width=0.48\columnwidth]{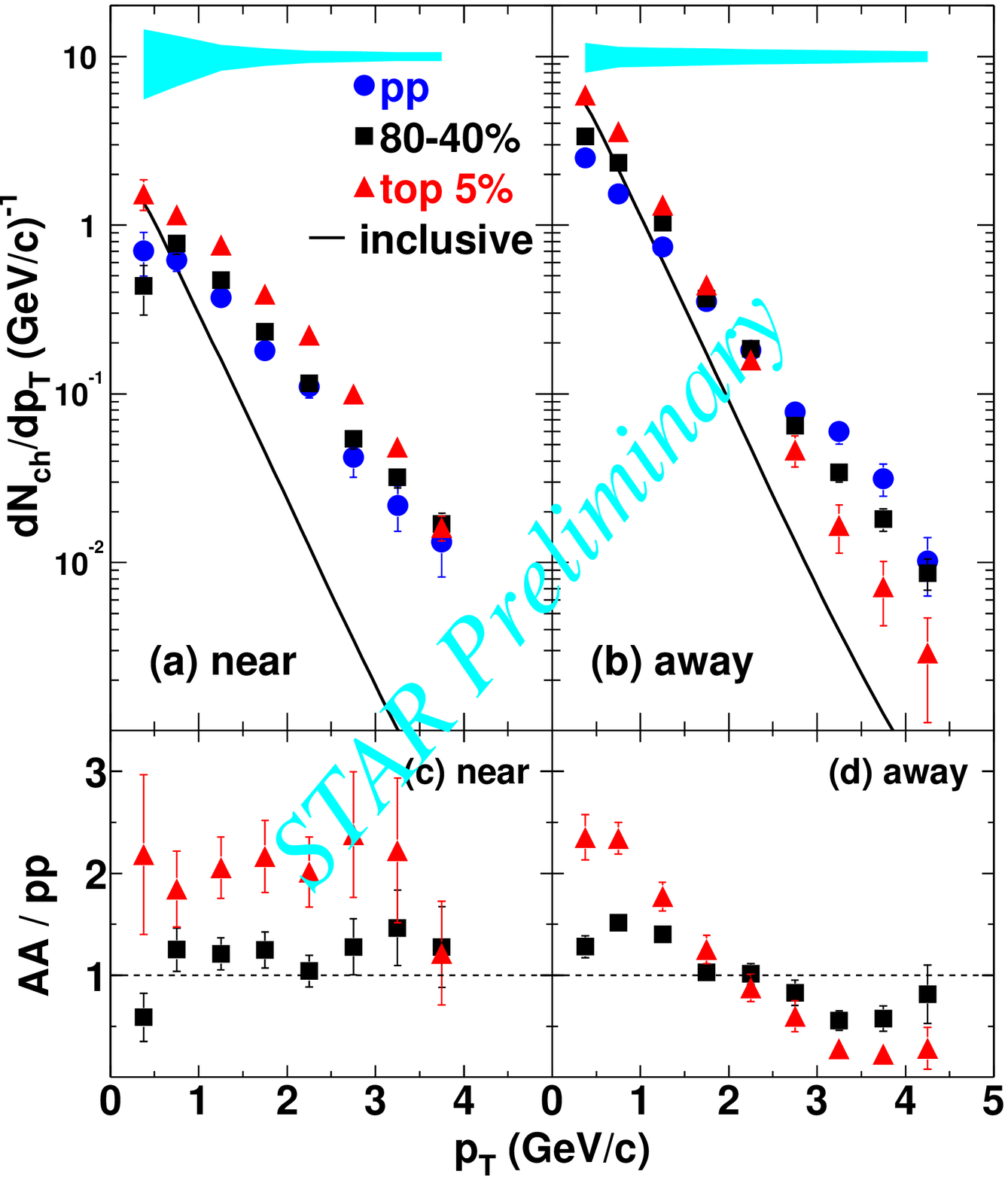}}\end{flushleft}
\caption{Left: $\Delta\phi$ distribution for 5\% central Au+Au collisions
measured by STAR \cite{fuqiang}. 
The solid line shows an estimate from the statistical momentum conservation
for the away side.
Right: STAR data \cite{fuqiang} 
on near-side (a) and away-side (b) \pt~distributions of associated charged hadrons for p+p, peripheral and central Au+Au collisions. Lower panels:
Ratios of Au+Au to p+p distributions. 
\label{fig:fuqi}}
\end{center}
\vspace{-0.5cm}
\end{figure}

Measurements of the redistribution of the energy
radiated off high energy partons in the medium would provide a very important
cross-check of jet quenching and, perhaps, a means
to probe the degree of thermalization.
First attempt of such studies was presented by STAR \cite{guo,fuqiang}. 
The analysis involves statistical reconstruction 
of charged hadrons associated with a high \pt~trigger particle
($p_T^{\rm trig}=$4-6 GeV/c) in azimuth and pseudorapidity.
The azimuthal distribution of the associated particles with 
\pt=0.15-4.0 GeV/c with respect to the trigger hadron, after
subtraction of the combinatorial background, exhibits jet-like
near-angle correlations in p+p and Au+Au collisions 
(Fig.~\ref{fig:fuqi}, left). On the away side, the distribution is
broad and consistent with a $\cos{\Delta\phi}$ shape, as expected for
statistical momentum balance \cite{borghini} with no additional 
dynamical correlations. The dependence of the correlation
magnitude on $p_T^{\rm trig}$ for central Au+Au collisions 
seems to be in agreement with the estimates
from momentum conservation, even for $p_T^{\rm trig}>$6.5 GeV/c \cite{guo}.
This may signal
equilibration of the medium-induced gluon radiation with 
momentum distributed over many particles. 

The correlation signals were integrated, and
charged hadron multiplicities and \pt~ distributions of associated particles
within $|\Delta\phi|<$1.0 and $|\Delta\eta|<$1.4 (near-side cone) and
$|\Delta\phi|>$1.0 (away-side cone) of the trigger hadron 
were studied as a function of centrality in Au+Au collisions and
compared to p+p collisions. Figure~\ref{fig:fuqi} shows the \pt~distributions of
associated particles. The away side is observed to be depleted
at large \pt~ and enhanced at low \pt~ compared to p+p collisions.
The broad angular distribution and softening of the spectrum 
may be consistent with predictions of 
the softening of the hadron-triggered fragmentation
function due to parton energy loss 
and enhancement of soft hadrons from emitted
gluons \cite{softening}.

\section{Conclusions}

To summarize, in the last year experiments have gained solid ground on 
elucidating jet properties at RHIC. Preliminary results on measurements 
of jet fragmentation in p+p and d+Au collisions provide a baseline
for comparison to Au+Au collisions. 
First direct evidence that back-to-back dijet quenching depends on the 
azimuthal orientation of the jets relative to the reaction plane
has been presented. Future high-statistics measurements of 
flavor-tagged di-hadron correlations will probe
energy loss effects of quark versus 
gluon jets and test string fragmentation versus recombination.
Finding the lost remnants of jets may provide a
cross-check of the picture of medium induced final-state energy loss.
Statistical reconstruction of jets in Au+Au collisions
has shown potential to 
get an experimental handle on the degree of 
thermalization at the partonic level.
In general, presented data broadly support pQCD-based models 
of final-state parton energy loss in a dense QCD medium 
produced in Au+Au collisions at RHIC and 
put strong experimental constraints on its properties.

This work was supported by the Director, Office of Science, Nuclear Physics, 
U.S. Department of Energy under Contract DE-AC03-76SF00098.


\section*{References}
\begin{thebibliography}{99}

\bibitem{quenching}
Gyulassy M and Pl\"{u}mer M 1990 
{\it Phys.~Lett.} B {\bf 243} 432 
\item[]Wang X N and Gyulassy M 1992
{\it Phys.~Rev.~Lett.} {\bf 68} 1480 
\item[]Baier R, Schiff D and Zakharov B G 2000
{\it Ann.\ Rev.\ Nucl.\ Part.\ Sci.}  {\bf 50} 37 
\bibitem{suppression}
Adcox K \etal (PHENIX Collaboration) 2002
{\it Phys.\ Rev.\ Lett.} {\bf 88} 022301
\item[]Adler C \etal (STAR Collaboration) 2002
{\it Phys.\ Rev.\ Lett.} {\bf 89} 202301
\item[]Adler S S \etal (PHENIX Collaboration) 2003
{\it Phys.\ Rev.\ Lett.} {\bf 91} 072301
\item[]Adams J \etal (STAR Collaboration) 2003
{\it Phys.\ Rev.\ Lett.} {\bf 91} 172302

\bibitem{tomography}
Gyulassy M, Levai P and Vitev I 2002
{\it Phys.\ Lett.} B {\bf 538} 282 
\item[] Wang E and Wang X N 2002
{\it Phys.\ Rev.\ Lett.} {\bf 89} 162301
\item[] Salgado C A and Wiedermann U A 2002
{\it Phys.\ Rev.\ Lett.} {\bf 89} 092303
\item[] Vitev I and Gyulassy M 2002
{\it Phys.\ Rev.\ Lett.}  {\bf 89} 252301

\bibitem{flow}
Adler C \etal (STAR Collaboration) 2003
{\it Phys.\ Rev.\ Lett.} {\bf 90} 032301

\bibitem{btob}
Adler C \etal (STAR Collaboration) 2003
{\it Phys.\ Rev.\ Lett.} {\bf 90} 082302

\bibitem{qm02}
{\it Proc. of the 16th Int. Conf. on Ultra-Relativistic Nucleus-Nucleus Collisions (Quark Matter 02)} 

\bibitem{dAuPhobos}
Back B B \etal (PHOBOS Collaboration) 2003
{\it Phys.\ Rev.\ Lett.} {\bf 91} 072302
\bibitem{dAuPhenix}
Adler S S \etal (PHENIX Collaboration) 2003
{\it Phys.\ Rev.\ Lett.} {\bf 91} 072303
\bibitem{dAuStar}
Adams J \etal (STAR Collaboration) 2003
{\it Phys.\ Rev.\ Lett.} {\bf 91} 072304
\bibitem{dAuBrahms}
Arsene I \etal (BRAHMS Collaboration) 2003
{\it Phys.\ Rev.\ Lett.} {\bf 91} 072305
\bibitem{denterria}
d'Enterria D 2004 these proceedings
\bibitem{miller}
Miller M 2004 these proceedings
\bibitem{vitev}
Vitev I 2004 these proceedings
\bibitem{baryonmeson}
Adler S S \etal  (PHENIX Collaboration) 2003
{\it Phys.\ Rev.\ Lett.}  {\bf 91} 172301 
\bibitem{henry} 
Henry T (STAR Collaboration) 2004 these proceedings
\bibitem{rak}
Rak J (PHENIX Collaboration) 2004 these proceedings
\bibitem{apanasevich}
Apanasevich L \etal 1999
{\it Phys. Rev.} D {\bf 59} 074007
\bibitem{averbeck}
Averbeck R 2004 these proceedings
\bibitem{bosing}
Klein-B\"osing C (PHENIX Collaboration) 2004 these proceedings
\bibitem{matathias}
Matathias F (PHENIX Collaboration) 2004 these proceedings
\bibitem{dAuStarId}
Adams J \etal  (STAR Collaboration) 2003
nucl-ex/0309012.
\bibitem{lamont}
Lamont M (STAR Collaboration) 2004 these proceedings
\bibitem{idflowphenix}
Adler S S \etal  (PHENIX Collaboration) 2003
{\it Phys.\ Rev.\ Lett.}  {\bf 91} 182301
\bibitem{idflowstar}
Adams J \etal  (STAR Collaboration) 2004
{\it Phys.\ Rev.\ Lett.} {\bf 92} 052302 
\bibitem{fries}
Fries R J 2004 these proceedings
\bibitem{sickles}
Sickles A (PHENIX Collaboration) 2004 these proceedings
\bibitem{guo}
Guo Y (STAR Collaboration) 2004 poster presentation, hep-ex/0403018
\bibitem{aihong} 
Tang A (STAR Collaboration) 2004 these proceedings
\bibitem{inout}
Bielcikova J \etal 2004
{\it Phys. Rev.} C {\bf 69} 021901(R)
\bibitem{wang}
Wang X N 2001
{\it Phys.~Rev.} C {\bf 63} 054902 
\item[]Gyulassy M, Vitev I and Wang X N 2001
{\it Phys.~Rev.~Lett.} {\bf 86} 2537 
\bibitem{kaneta} 
Kaneta M (PHENIX Collaboration) 2004 these proceedings
\bibitem{shuryak}
Shuryak E V 2002
{\it Phys.\ Rev.} C {\bf 66} 027902
\bibitem{fuqiang} 
Wang F (STAR Collaboration) 2004 these proceedings
\bibitem{borghini}
Borghini N, Dihn P M and Ollitrault J Y 2000
{\it Phys. Rev.} C {\bf 62} 034902
\bibitem{softening}
Wang X N 2004 
{\it Phys.Lett.} B {\bf 579} 299 
\end{thebibliography}
\end{document}